\documentstyle[11pt,revpasp,epsf]{article}
\begin{document}

\title{Nature and evolution of Damped Lyman alpha systems}
\author{G. Vladilo}
\affil{Osservatorio Astronomico di Trieste, Via Tiepolo 11,
34131 Trieste, Italy}

\begin{abstract}
The main properties of Damped Lyman alpha (DLA) systems 
are briefly reviewed with the aim of studying 
the nature and evolution of the galaxies associated with this
class of QSO absorbers.  
Candidate DLA galaxies identified at $z \leq 1 $
in the fields of background QSOs show a variety of 
morphological types without a predominance of spirals. 
Most properties inferred from
spectroscopic studies at $z \geq 1.65$
differ from those expected for spiral galaxies. 
The observational results instead suggest that
a significant fraction of DLA systems originate 
in low-mass and/or LSB galaxies. 
Evolution effects are generally not detected in DLA systems.
This fact suggests that the
differences between the properties of present-day spirals
and those of high-$z$ DLA systems
may not be ascribed to evolution.    
Several selection effects can bias the observed
population of DLA absorbers.  Analysis of these effects indicates
that the fraction of spiral galaxies tends to be underestimated
relative to the fraction of low-mass or LSB galaxies. 
\end{abstract}

\section{Introduction}
 
The redshifted Ly$\alpha$ absorptions 
observed in the spectra of QSOs originate in intervening clouds with
wide-ranging HI column densities. 
When  $N$(HI) $\geq$ 2 $\times$ 10$^{20}$ atoms cm$^{-2}$
the Ly $\alpha$ line shows a broad profile
with extended 'radiation damping' wings. 
Damped Ly $\alpha$ (DLA) absorption lines are always 
accompanied by
narrow metal lines at the same redshift, $z_{\rm abs}$.
These absorption line systems are quite rare, 
the number per unit redshift interval being the
lowest among all types of QSO absorbers
($n(z) \simeq 0.2$ at $z_{\rm abs} \simeq 2$;
Wolfe et al. 1995). 
About a hundred DLA systems are currently known
as a result of several surveys, most of them performed in the optical
spectral range (Wolfe et al. 1986, 1995). 
Owing to the dramatic drop of QSO counts
at high $z$, only a reduced number of
absorbers have been detected at $z>4$ 
(Storrie-Lombardi et al. 1996a).
At $z \leq 1.65$   Ly $\alpha$ absorptions  
can only be observed with space-born UV telescopes
and the number of systems identified is quite limited
(Lanzetta, Wolfe \& Turnshek 1995; Jannuzi et al. 1998
and refs. therein; Turnshek 1998). 
DLA systems have also been detected  
as redshifted 21 cm absorption in the continuum of radio
loud quasars (Carilli et al. 1996 and refs. therein). 
 
Several reasons suggest that  
DLA systems originate in interstellar clouds within galaxies
located in the direction of the QSO:
(i) the high values of HI column density,  typical of the 
interstellar medium of gas-rich galaxies;
(ii) the presence of low ionization species of metals, observed
in Galactic HI regions;
(iii) the line-of-sight  velocity dispersion,  
consistent with the typical values expected for   galactic disks;
(iv) the evolution of
the comoving mass density of gas in DLA absorbers,  which is suggestive of
gas consumption due to star formation
(Wolfe et al. 1995). 
DLA systems have low metallicities, typically
$Z/Z_\odot \approx  10^{-1}$ and, in some cases,
as low as $Z/Z_\odot \approx$ 10$^{-2}$  
(Pettini et al. 1997a, 1999). 
Therefore the galaxies hosting DLA clouds 
must be chemically young and,
in some cases, must be in the very
first stages of their chemical enrichment.

Studies of DLA absorbers allow us 
to probe young galaxies at high redshift 
from the observation of only one  line of sight
though each galaxy.  
This kind of investigation is complementary
to studies of Ly-break galaxies, 
which allow us to probe high redshift galaxies
from the observation of their integrated emission. 
The advantage of DLA absorption studies
is the intrinsic brightness of the background QSO
which can be used to obtain spectra
of unparalleled resolution and
signal-to-noise ratio at a given redshift. 
The quality of these spectra  allows us to
study the chemical and physical
properties of young galaxies in unrivaled detail. 
 
Even if the link between DLA absorbers and intervening galaxies
is commonly accepted, there is no general agreement 
on the nature of the galaxies hosting the DLA clouds
(also called DLA galaxies hereafter). 
The traditional working hypothesis is that
DLA galaxies are the progenitors of present-day spirals
(Wolfe et al. 1986), but alternative interpretations have
also been proposed (e.g. Tyson 1988). 
While it is possible to study the evolution of DLA systems
{\it per se}, understanding 
the nature of DLA galaxies is fundamental 
to put the phenomenon in the
general context of galactic evolution 
and to constrain theories of structure formation. 
The observational clues to the nature of DLA galaxies
are summarized
in the next section of this contribution. 
Selection effects  are considered in Section 3,  while 
the evolution properties are discussed in 
Section 4. Finally, the results are summarized in Section 5.

\section{ Clues to the nature of DLA galaxies }

One approach to cast light on the nature of DLA galaxies
is to estimate the number of galaxies
of a specific morphological type $T$ expected along
a random line of sight. 
At   $z\simeq 0$ the number per unit distance interval is
\begin{equation}
n^T_\circ = 
\int \Phi^T_\circ (M) \, <A^T_\circ(M)> dM 
\end{equation}
where
$\Phi^T_\circ$ is the optical luminosity function of the  
galaxies of   type $T$  in terms of the absolute magnitude $M$;
$A^T_\circ$ is the effective cross section of the column density
contour   $N_{\rm HI} \geq N_{\rm min} =
2 \times 10^{20}$ cm$^{-2}$
and the angled brackets indicate a weighted average over all possible
galaxian inclinations. 
Estimates of the HI content within galaxies at the present epoch 
indicate that most of the observed absorptions  should
originate in spirals (Rao \& Briggs 1993).
This prediction, however, is
not confirmed by studies of  candidate DLA galaxies
at low redshift (Section 2.1). 
Considering the lack of understaning at $z=0$ it is clear that
estimating the fraction $n^T_\circ/ \sum_T n^T_\circ$
at high $z$ is highly speculative 
until we know the effects of evolution and merging
on morphology and galaxian sizes.

Observations are the key to understanding the nature of 
DLA galaxies.  
Spectroscopic data are used to study chemical and physical properties.
Imaging data are used to study  
the morphology of candidate DLA galaxies. Imaging and
spectroscopic data have a poor redshift overlap:
while the imaging is most effective at $z \leq 1$ ---
the confusion with the QSO source is more critical at high $z$ ---
the spectroscopy is mostly performed at $z_{\rm abs} \geq 1.65$.

\begin{table}
\caption{ Summary of searches for candidate DLA galaxies }
\begin{tabular}{llllclc}
\tableline
QSO & ~~$z_{\rm abs}$ &  ~~$z_{\rm gal}$ & ~~$\rho^a$ & $M_B$ or & Type & Ref. \\
    &                 &                & (kpc)  & Luminosity & &      \\
\hline
Ton 1480 & 0.0039  & 0.0039 & 12 & ---  & S0  &  1 \\
OI 363 & 0.0912 &            & --- & $\geq$--15.9$^b$ & [dw/LSB] & 2 \\
0850+4400 & 0.1638 & 0.1635 & 33 & $L_B = 0.4 \, L^*_B $& S0  & 3 \\
OI 363  &  0.2212  &     &  ---   & $\geq$--17.8$^b$ & [dw/LSB] & 2 \\
PKS B1127--145 & 0.3127 & 0.3121 & 22 & $m_R = 22.3$ & --- & 4 \\
1229--021 & 0.3950 &    & 9.9    & --18.9 &  LSB  & 5 \\
3C 196 & 0.437  &    & 12.5   & -22.1 & Giant Sbc  & 5 \\
PKS 0118--272 & 0.5580 &   &  14  & $\approx-21.3^b$ & --- &  6  \\
1209+107 & 0.6295 &  & 14.6   & --22.0 & Spiral     & 5 \\
3C 336  & 0.6563   &  & ---   & $L < 0.05 \, L^*_K $ & [dw/LSB] & 7 \\
1328+307 & 0.692  &  & 8.5   & --20.5 & LSB         & 5 \\
1331+170 & 0.7443 &  & 37.7   & -22.9 & Spiral & 5 \\
0454+039 & 0.8596 &  &  8.3  & --20.5 & Compact  & 5 \\
0302--223 & 1.0095 &  & 12.0   & --20.4 & Semicomp.  & 5 \\
          &        &  & 27.4   & --22.0 & Compact  & 5 \\
1331+170 & 1.776  &   & 20.0   & --22.7 & Compact  & 5 \\
\tableline\tableline
\end{tabular}
\par\noindent
$^a$ All impact parameter have been converted
in units of $h_{50}^{-1}$ kpc. 
\par\noindent
$^b$ Derived by assuming $B-R=1$ (Rao \& Turnshek 1998)
\par\noindent
REFERENCES.--- 
(1) Miller, Knezek, \& Bregman 1999; (2) Rao \& Turnshek 1998;
(3) Lanzetta et al. 1997; (4) Lane et al. 1998;
(5) Le Brun et al. 1997; (6) Vladilo et al. 1997;
(7) Steidel et al. 1997. 
\end{table}

\subsection{ Morphology of candidate DLA galaxies  }

The galaxies responsible for
the DLA absorption can  be identified by studying
the field of the background QSO.  
Galaxies with impact parameter compatible with 
the expected extension of  the HI disk  
are considered as candidate absorbers.
The impact parameter, $\rho$ ($h_{\circ}^{-1}$ kpc),
is estimated at the redshift of the absorber and, in order
to confirm the
identification, one should also measure the redshift of the galaxy
and check if $z_{\rm gal} = z_{\rm abs} $.
When galaxies are not detected within a reasonable 
value of $\rho$,
an upper limit to the (surface) brightness of the intervening galaxy
can   still be derived. 
The results of
searches for DLA galaxies in QSO fields   
are summarized in Table 1.  
Although the sample is limited and
only a few   galaxies have a redshift measurement, 
an important conclusion can already be derived:
DLA galaxies at $z \leq 1$ show a variety of 
morphological types (S0, spirals, dwarfs) and 
different levels of surface brightness,
including low surface brightness (LSB) galaxies.
In other words, the population of DLA galaxies is not
dominated by any specific type of galaxies and, in particular,
spirals  constitute a small fraction of the sample,
contrary to the predictions based on the HI content
of nearby galaxies (Rao \& Briggs 1993). 
The selection effects discussed in Section 3 may be responsible
for this unexpected result.

\subsection{ Elemental abundances   }
 
Abundances of DLA systems can be measured with  accuracy  
and are already available for about 50  systems 
(Lu et al. 1996; Prochaska \& Wolfe 1999;
Pettini et al. 1999;  see also
refs. in Vladilo 1998). 
The HI column density can be easily constrained within $\pm$ 0.1 dex,
or even better, 
by fitting the damping wings of the Ly $\alpha$ profile. 
The most common metals   show unsaturated transitions 
which allow column densities to be accurately determined.   
Ionization corrections are generally negligible thanks to the presence
of neutrals or ions with IP $\geq$ 13.6 eV
which are dominant ionization stages in HI regions. 
Dust probably represents
the main source of uncertainty in abundance determinations
since an unknown fraction of the elements is probably
depleted into dust grains (Section 2.3).
Studies of the  intrinsic abundances of DLA systems  
in the presence of dust  have been performed by
Lauroesch et al. (1996) and by Kulkarni, Fall \& Truran (1997). 
In these studies the  
dust-to-gas ratio, $k$, is considered a 
free parameter with same value in all systems. 
However,   the level of depletion   scales with  $k$, and
it is essential to estimate $k$ for each DLA cloud in order
to properly correct  the  abundances (Vladilo 1998). 

Abundances of metal-poor Galactic stars are often
used as a reference for DLA studies since they  
reflect the abundance pattern of the Milky-Way   gas  
at the time in which the first stellar generations were formed. 
By comparing abundances of DLA systems at redshift $z$ with
abundances of Galactic stars formed at look-back  
time $ t(z)$, we can test whether DLA galaxies
undergo a chemical evolution similar to that of the Milky Way.
The comparison between the two sets of abundances
can also be made at a given metallicity, which measures the level
of chemical enrichment attained by each system. 

\subsubsection{Metallicities.}
The absolute abundance of zinc, [Zn/H]\footnote
{ We adopt the usual notation
[X/Y] $\equiv$ log\,(X/Y) -- log\,(X/Y)$_{\odot}$ },
is generally used to study the 
metallicities in DLA systems since zinc is expected to be unaffected
by dust depletion (Pettini et al. 1997a; 1999). 
Observed metallicities span
the interval $ -2 < {\rm [Zn/H] } < 0$, with a column-density weighted 
mean value [$<$Zn/H$>$] $\simeq -1 $. 
The metallicity distribution
is different from that found in the stellar
populations of the Milky Way, a result that casts 
doubts on the relationship between high-$z$ DLA systems 
and present-day spirals 
(Pettini et al. 1997b; see, however, Wolfe \& Prochaska 1998). 
 
\subsubsection{ Iron-peak abundances. }
Studies of metal-poor stars in the Galaxy indicate that
iron-peak elements trace each other
with approximate solar ratios  
(Ryan, Norris, \& Beers 1996).
Deviations from the solar pattern can be present, but are generally 
negligible at the   metallicity level of
DLA absorbers.
The [Zn/Fe], [Cr/Fe],  and [Ni/Fe] ratios 
measured in DLA systems show  significant deviations 
from the solar pattern, inconsistent with those observed
in metal-poor stars
(Lu et al. 1996; Pettini et al. 1997a). 
All these ratios follow the differential dust depletion
pattern observed in the Milky Way interstellar gas
(Savage \& Sembach 1996), suggesting that the observed
abundances are dominated by dust depletion 
(Section 2.3). 
The [Mn/Fe] ratio is underabundant, consistent with that observed
in metal-poor stars (Lu et al. 1996); however   part of this effect
can also be ascribed to dust  (Vladilo 1998).

\begin{figure}
\plotfiddle{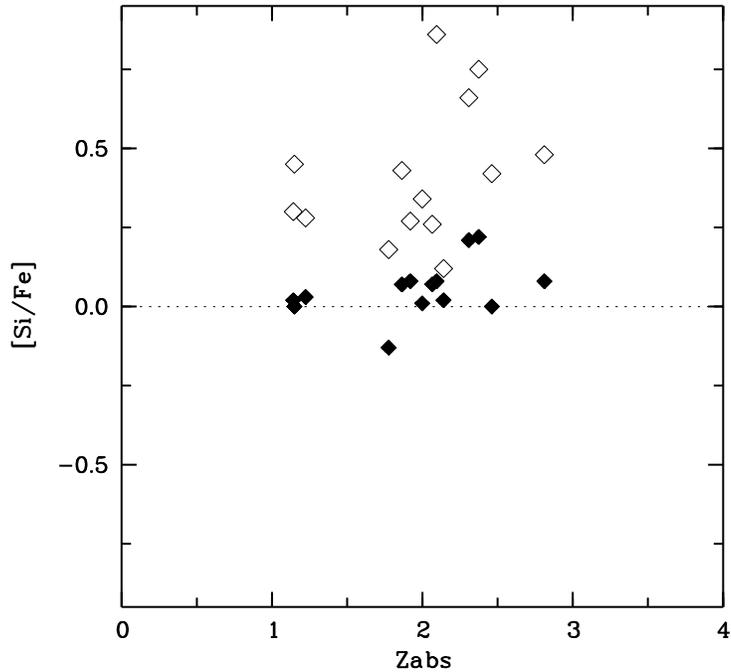}{8cm}{0}{80}{80}{-240}{-260}
\caption{
Empty symbols:   measurements of the [Si/Fe] ratio in DLA systems. 
Filled symbols: same measurements corrected for dust depletion. 
See Vladilo (1998) for more details. 
}
\end{figure}

\subsubsection{ [$\alpha$/Fe] ratios.}
The ratio between $\alpha$ and iron-peak elements 
is a well-known tracer of galactic evolution. In the Milky Way
it decreases  
from the value [$\alpha$/Fe] $\simeq$ +0.5 typical
of  metal-poor stars, down to [$\alpha$/Fe] = 0 
at higher metallicities 
(Wheeler, Sneden \& Truran 1989). 
The temporal delay between the metal injection from SNae Type II,
rich in $\alpha$ elements, and   SNae Type Ia,
rich in iron-peak elements 
can explain the decrease of  [$\alpha$/Fe] in the course
of evolution
(Matteucci 1991 and refs. therein). 
The [Si/Fe] and [S/Fe] ratios in DLA systems
show overabundances which resemble
Milky-Way metal-poor abundances  (Lu et al. 1996).
However,   the result in itself is not conclusive since 
the [Si/Fe] and [S/Fe] ratios are also enhanced    
in the nearby ISM as a consequence
of differential depletion (Savage \& Sembach 1996). 
A way to obtain the intrinsic [$\alpha$/Fe] ratio
is to select elements with negligible ISM depletions,
such as sulphur and zinc, which trace  the
$\alpha$   and  the iron-peak elements, respectively.
The few available measurements give  [S/Zn] $\simeq$ 0 
(Molaro, Centuri\'on, \& Vladilo 1998), 
suggesting that the enhancement of the 
[Si/Fe] and [S/Fe] ratios is   due to differential  depletion.
This suggestion is confirmed by a re-analysis
of abundances in DLA systems corrected for depletion
(Vladilo 1998): 
the resulting [Si/Fe] and [S/Fe] values are
approximately solar (Fig. 1), consistent with the [S/Zn] results. 
The Magellanic Clouds (Wheeler et al. 1989)
and  BCGs (Thuan, Izotov \& Lipovetsky 1995) are examples
of galaxies  with [$\alpha$/Fe] $\simeq$ 0 at low metallicity.
In general,  any
galaxy with low SFR at early epochs will be able to produce
[$\alpha$/Fe] $\simeq$ 0 at low metallicity   
because the onset of Type Ia SNae will
occur before the galaxy has time to attain   solar metallicity
(Matteucci 1991).

\subsubsection{ Nitrogen.}

Nitrogen abundances can be used to probe   the early
stages of chemical evolution.  However, 
the relative importance of different production mechanisms
--- i.e. primary versus secondary production ---
is not fully understood  
(Matteucci, Molaro, \& Vladilo 1997). 
Nitrogen abundances     have been measured in about ten DLA systems
(Molaro et al. 1996;
Lu, Sargent \& Barlow 1998; Centuri\'on et al. 1998).
When the effects of dust are considered,
a substantial fraction of [N/Fe] and [N/S] ratios
are lower than  those observed in Galactic metal-poor stars
(Centuri\'on et al. 1998).  
It is not possible to explain  all the observations with
a unique production mechanism: some cases suggest
a secondary behaviour 
(i.e. the nitrogen ratios increase with metallicity), 
whereas others show evidence of primary  production 
(i.e. the  ratios are approximately constant with metallicity).
Nitrogen ratios in DLA systems show  
similarities with those  
measured in nearby metal-poor galaxies, such as
the [N/O] ratios in dwarf irregulars (Kobulnicky \& Skillman 1996;
van Zee et al. 1996) and the [N/Fe] ratios in BCGs 
(Thuan et al. 1995). 
In one or two cases there is evidence for an extremely
high [N/$\alpha$] ratio, well above the values found
in any astrophysical site 
(Molaro et al. 1996).

\subsection{ Dust  }
 
The first evidence for dust in DLA systems was provided
by a study of QSOs with and without
foreground DLA absorption (Pei, Fall \& Bechtold 1991).  
The optical spectral indices of the two samples 
are significally different and indicate an enhanced
reddening  of the QSOs with intervening absorption. 
The dust-to-gas ratios, $k$,  derived from this statistical study  
are typically between 5\% and 20\% of the Galactic value. 

The observation of different images of   
gravitationally lensed QSOs with foreground
DLA absorption is a powerful technique to study the 
dust properties of the intervening galaxy.  
The only case investigated up to now, namely the
$z_{\rm abs}$ = 1.3911 system toward QSO 0957+561,
shows clear evidence of differential dust reddening 
between  the two adjacent images of the QSO
(Zuo et al. 1997). The derived  $k$
is between 40\% and 70\% of the Galactic value. 
 
As mentioned above, also the   abundances of 
iron-peak elements provide evidence for dust in DLA systems
since the [Zn/Fe], [Cr/Fe], and [Ni/Fe] ratios 
qualitatively follow the dust depletion pattern seen in 
the nearby interstellar gas (Savage \& Sembach 1996).  
The observed pattern can be quantitatively
explained by assuming that the dust in DLA systems 
has same the composition as in the Milky Way, but with a different  
value of $k$ in each system.  
In fact, one can estimate the dust-to-gas ratio in
each system from the condition that the intrinsic
iron-peak abundance ratios are solar (Vladilo 1998).  
The resulting dust-to-gas ratios show a large spread
among different DLA absorbers, with $k$ values
mostly distributed between 2\% and 25\% of the Galactic value,  
consistent with the range found by Pei et al. (1991).
In a given DLA system, 
dust-to-gas ratios estimated from different pairs
of iron-peak elements  yield  consistent results,
as expected from the basic assumption of the method. 
Dust-to-gas ratios estimated in this way 
are well correlated with metallicity  
(Fig. 2 in Vladilo 1998). 
The existence of such a correlation and the evidence
for dust obscuration described in Section 3.1 
indicate that the $k$ values estimated
indirectly from the iron-peak abundances  
are indeed related to dust present in DLA systems.

\subsection{ Kinematics  }

Kinematical properties are derived from the study
of the line-of-sight velocity dispersion  of the systems, 
determined from the   profiles of 
unsaturated metal lines.
The  HI gas is traced by low ions, such as
Si$^+$ or Fe$^+$, which   have very similar  profiles 
in each system and   often show   
multiple absorption components. 
Since the line of sight samples the absorbers  along random directions,
a large number of observations is required to test  
models of DLA kinematics. 
The most extensive collection of velocity profiles has been
obtained by Prochaska \& Wolfe (1997, 1998) by means of Keck
observations.  
The velocity widths, $\Delta V$,  measured above a given
threshold of optical depth, range  
from about 50 km s$^{-1}$ up to 300 km s$^{-1}$.
When multiple components are present, the
most intense one is generally found at one edge of
the profile. 
These "leading-edge" profiles can be naturally produced 
by the intersection of a rotating disk with
 an exponential gas density distribution. 
Analysis of the full set of profiles indicates a consistency
with models of fast-rotating 
($V_{\rm rot} \simeq 250$ km s$^{-1}$), thick disks,
a result supporting a relationship between high-$z$
DLA systems and present-day spirals; 
models of slow-rotating, low-mass   galaxies are instead ruled
out (Prochaska \& Wolfe 1997, 1998).
However, these conclusions are obtained by
assuming that all DLA systems are drawn from a
homogeneous population of galaxies, 
an assumption  probably
incorrect, given the observational evidence shown in Table 1.  
In addition, the conclusion that fast rotating disks are the only viable
explanation for the observed data has been disproved
by Haehnelt et al. (1998). According to these authors,
irregular protogalactic clumps can reproduce the  velocity
profiles distribution equally well. The few profiles with extremely high
values of $\Delta V$ can be explained with 
occasional alignment
of clumps at the same redshift. 
An argument against the hypothesis that DLA systems
rotate at
$V_{\rm rot} \simeq 250$ km s$^{-1}$ comes
from an analysis of profile asymmetries 
performed by Ledoux et al. (1998). According to these
authors,  there is evidence for regular rotating disks only up to
$\Delta V_{\rm rot} \simeq 120$ km s$^{-1}$, 
while at higher velocities the kinematics is more complex.

\subsection{ Spin temperature  }

For DLA systems that lie in front of a radio loud quasar
it is possible to observe the 21 cm absorption
in addition to the  Ly $\alpha$ line.  
An analysis of both spectral ranges
yields, under suitable assumptions,
the harmonic mean along the line of sight of the
spin temperature of the gas, $<T_s>$. 
The typical values found in DLA systems --- 
$<T_s> \approx 10^3$\,K  
(de Bruyn, O'Dea \& Baum 1996; Carilli et al. 1996) ---, 
are generally much larger
than those observed in the disk of the Galaxy or in
nearby spiral galaxies  
(Braun 1997 and refs. therein). 
The higher spin temperatures probably indicate that
DLA galaxies have a larger fraction of warm gas than
nearby spirals. 
One approach to understanding this difference is through  
variation of the interstellar pressure: 
the fraction of warm gas is expected to be
higher in regions where the mean pressure is lower (Dickey 1995).
Since the mean pressure is determined, in part, by the 
gravitational potential, a high fraction of warm gas
could be a signature of gravitational potential lower
than in the Milky Way disk.

\section{ Selection effects }

Selection effects can alter the fraction of specific types
of galaxies, or particular regions of galaxies,
detected in surveys of DLA systems. 
Recognizing the role of such effects is fundamental 
for a correct interpretation  of   
the nature and evolution of DLA galaxies. 

\subsection{ QSO obscuration  }

The absorbers with the highest dust content 
will obscure the background QSO and 
will be missed from magnitude limited  samples.
This effect of QSO extinction was first investigated by
Pei et al. (1991). 
The possibility of determining the dust-to-gas ratios  $k$
in individual systems allows us to estimate the importance
of the effect  
(Vladilo 1998).  
Evidence for QSO obscuration comes 
an inspection of Fig. 2: DLA systems for which 
the product ${\cal D}  = k N_{\rm HI}$
exceeds a critical threshold are not observed. 
${\cal D}$  is an estimate of the dust content along
the line of sight and
the tilted line in Fig. 2 represents the  ${\cal D}$ value
that yields an extinction
of 1 magnitude of the QSO in the observer's frame. 
Absorbers above this line have
probably not been detected because they
obscure the QSO by more than 1 magnitude. 
Since dust and metals are strinctly linked, one  
expects that systems with  high metallicity and
high column density  
are missed due to the same selection bias.
This effect has indeed been reported by Boiss\`e et al. (1998). 

As a consequence of  QSO obscuration, DLA absorbers
with higher and higher dust content (or metallicity)
are only detectable at lower and lower values of column density. 
In particular, present-day spirals with solar metallicity 
(and hence $k/k_{\rm Gal} \approx 1$)
can be missed when 
$N_{\rm HI} \geq 10^{20.7}$ cm$^{-2}$, 
according to the trend shown in Fig. 2. 
Dwarf or LSB galaxies, which are
characterized by lower metallicities and dust content,
should be less affected by this selection bias.  
In addition, LSB galaxies should be less affected because 
the   column density perpendicular to the disk is typically
lower than in high surface brightness (HSB) galaxies.

\subsection{ Surface brightness }
 
Differences in surface brightness between galaxies
can be understood if 
LSB galaxies are hosted in dark halos with values of the spin parameter,
$\lambda$, larger than those of HSB galaxies
(see Jimenez, Bowen \& Matteucci 1999 and refs. therein). 
The cross-sections of individual galactic disks 
in equilibrium  scale as $\lambda^2$ and therefore 
will be dominated by objects with large angular momentum
(Mo, Mao \& White 1999). 
As a result, 
LSB galaxies are expected to dominate the cross-section for DLA absorption. 
On the other hand, HSB galaxies are expected to dominate the rates
of star formation and metal production.

\subsection{ Galactocentric distance }

The probability of detecting a galaxy
in the interval of galactocentric distances [$r, r+dr$]
is, in general, a function of $r$.
For a galactic disk seen face on, 
the differential cross section for DLA absorption
is $d A_\circ \simeq
2 \pi r \, dr$, 
until   $ N_{\rm HI} \geq N_{\rm min} $.
Therefore, galactic regions with larger $r$   have a higher
probability of detection, 
unless the galaxy is seen exactly edge on. 
Any property which shows spatial gradients
will be affected by this bias.
In particular, our understanding
of chemical evolution properties of DLA systems
will be biased since external regions are less chemically
evolved than inner regions.

\begin{figure}
\plotfiddle{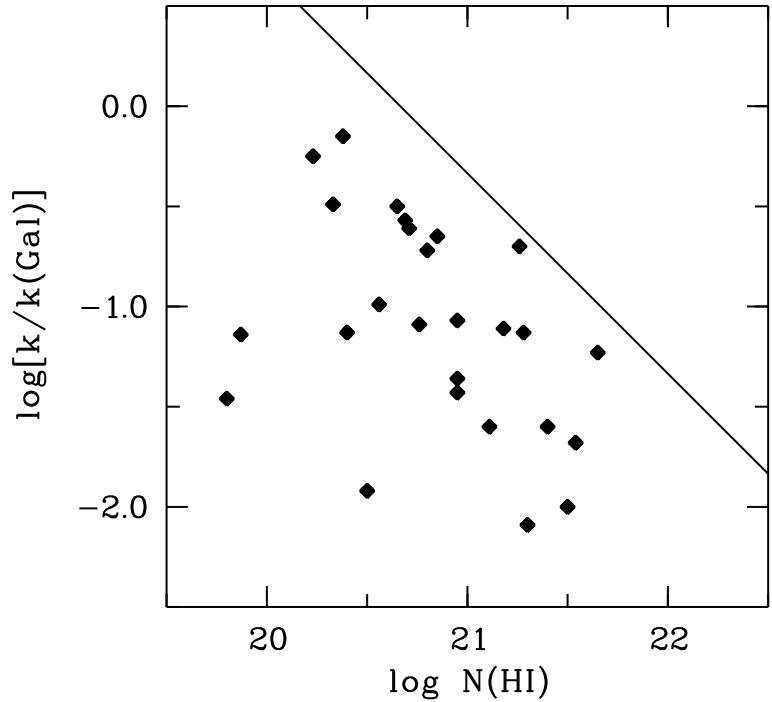}{8cm}{0}{80}{80}{-240}{-260}
\caption{
Dust-to-gas ratios (in units of the Galactic value)
versus HI column density in DLA systems.
Tilted line:   line of costant extinction  
of the  QSO (1 magnitude  in the observer's frame V band)
for an adopted SMC extinction curve 
(Vladilo, Molaro \& Centuri\'on 1999). 
}
\end{figure}

\subsection{ Gravitational lensing }

The   galaxy hosting the DLA system can act 
as a gravitational lens on the image of the background QSO. 
Smette, Claeskens \& Surdej (1997) have developed a formalism
to compute the effects  
of gravitational lensing ('by-pass' effect and 'amplification bias') 
on the observed number density of DLA systems. 
The 'by-pass' effect causes the line of sight to avoid the central
part of the intervening galaxy and to decrease its effective cross-section
for   absorption. The 'amplification bias' 
boosts the apparent magnitude of the QSO and therefore 
increases the fraction of QSOs with foreground galaxies in 
magnitude-limited samples. The 'amplification bias' 
acts in the opposite direction of dust obscuration. 
However, in order to predict the overall effect one should 
model dust obscuration and gravitational lensing 
in a self-consistent way. It is interesting to note that
both the 'by-pass' effect, and the
'galactocentric distance' bias
conspire   to exclude from the surveys
the inner regions of galaxies,  i.e. the most chemically evolved regions.

\section{ Evolution of DLA systems } 

Detecting evolution effects in DLA systems is difficult
for several reasons. 
First, the low number of absorbers identified
at low redshift represents a severe limitation, since 
$z \leq 1.65$ corresponds to a look-back time of about 
2/3 of the age of the universe.
Second, the selection effects mentioned in the previous
section imply that the observed samples are biased. 
Third, the samples are not homogeneous, since they
can include
galaxies of different morphological types $T$,
absolute magnitudes $M$, and  redshifts of formation $z_f$;
moreover  
the line of sight   crosses the galaxy
at a random  radius $r$. 
Therefore, observations of any physical
quantity $Q$ at different redshifts $z_{\rm abs}$
will yield  a data set of the type
$ Q = Q(z_{\rm abs}; T, M, z_f ; r)$, where the redshift
dependence of $Q$ will be disguised by fluctuations induced by
the other variables.

\subsection{ Number density and comoving mass density }

The number of absorbers per
unit redshift interval   is given by the product
of the absorber cross section, $A_\circ$,
times the number density of absorbers per comoving volume,
$\Phi_\circ$. 
The expression for a  standard Friedmann universe,
$n(z) = \Phi_\circ \, A_\circ \, c \, H_\circ^{-1} \,
 (1+z) \,  (1+2q_\circ z)^{-1/2}$,
is usually replaced  with 
\begin{equation}
n(z) = n_\circ (1+z)^\gamma  \, ,
\end{equation} 
where $\gamma$ is determined from the
best fit to the  empirical data points. 
In absence of intrinsic evolution $\gamma$ =
1/2 ($q_\circ=0.5$) or $\gamma$ = 1  ($q_\circ=0.$). 
The work by Lanzetta et al. (1995)
yields $\gamma=1.15 \pm 0.55$, consistent 
with no evolution.  
From a combined sample including DLA systems
at  $z \geq 4$, Storrie-Lombardi, Irwin \& McMahon (1996b) 
find $\gamma=1.3 \pm 0.5$,
also consistent with no evolution. 
However, these authors
do find evolution for the absorbers with 
$N_{\rm H I} > 10^{21}$ cm$^{-2}$, which show a
decline at $z\geq3.5$. 
By combining the sample of DLA systems
with the constraints on present-day galaxies, 
Rao, Turnshek \& Briggs (1995) find $\gamma= 2.27 \pm 0.25$.
However, this result should not be taken as evidence for evolution
until we understand clearly the link
between DLA absorbers and nearby galaxies.

The mean comoving mass density of gas in DLA systems 
is given by
\begin{equation}
 \Omega_{\rm dla}(z)  =
{ H_\circ \, \mu \, m_{\rm H} \over c \, \rho_{\rm crit} }
\int_{N_{\rm min}}^{\infty} N \, f(N,z) \, dN \, , 
\end{equation}
where 
$\mu$ is the mean molecular weight of the gas,
$\rho_{\rm crit}$ is the current critical density of the universe
and $f(N,z)$ is the column density distribution function
(see Storrie-Lombardi, McMahon, \& Irwin 1996c and refs. therein). 
Analysis of a sample including systems
at $z > 4$ indicates that 
$\Omega_{\rm dla}(z) $ increases from 
$z=0$ to $z\approx 2.5$ and apparently declines 
at $z > 3.5$  (Storrie-Lombardi et al. 1996c).  
In the lowest redshift bin 
$\Omega_{\rm dla}(z) $  is roughly equal to the comoving density
of neutral gas 
derived from 21-cm emission surveys of nearby galaxies, i.e.
$\Omega_{\rm dla}(z \simeq 0.64) \approx \Omega_{\rm 21 cm}(0) $.
At the peak value, $\Omega_{\rm dla}(z) $
is marginally consistent with  
the mass density in stars in nearby galaxies, i.e. 
$\Omega_{\rm dla}(z \approx 2.5) \leq \Omega_{\rm star}(0) $.
These two facts suggest that the evolution of 
$\Omega_{\rm dla}(z)$ from the peak value to the present-day value
is governed by gas consumption due to star formation,
as suggested by Wolfe et al. (1995). 
However,  if the inequality 
$ \Omega_{\rm dla}(z) < \Omega_{\rm star}(0) $ holds true,
we are missing  part of the gas responsible for the 
formation of present-day stars. This could be a consequence of
the QSO obscuration bias,  which  
affects the absorbers with highest column densities, i.e.
the absorbers that  give a dominant contribution
to $\Omega_{\rm dla}(z)$ (Eq.3).

A new study of DLA systems at low redshift suggests that
$\Omega_{\rm dla}(z)$  could be roughly constant
from $z\simeq 0.4$ up to $z\simeq 3$, with
$  \Omega_{\rm dla}(z) > \Omega_{\rm 21 cm}(0) $ 
(Turnshek 1998). This finding  would imply that  
the bulk of star formation took place only 
relatively recently. However, the result must be considered
with caution, since it is based on a very low number
of systems and may  also be affected by the 
gravitational lensing   bias  (Turnshek 1998). 

Jimenez, Bowen \& Matteucci (1999) have recently computed
model predictions of $\Omega_{\rm dla}(z)$
for galaxies with low and high levels of surface brightness. 
According to these authors, HSB galaxies consume neutral
gas at a rate which is too fast to explain
the observed evolution of $\Omega_{\rm dla}(z)$; instead
LSB galaxies provide 
a good fit to the data  published by Storrie-Lombardi et al. (1996c). 

\subsection{ Metallicity and abundance ratios } 

Metallicities are expected to increase in the
course of evolution. 
Measurements of  the [Zn/H] ratio 
are now available for 40 DLA systems, including 10 absorbers at $z<1.5$
(Pettini et al. 1999).  The analysis of this sample
does not reveal evidence for an increase with time: 
the column-density weighted mean metallicity 
is not significantly higher at  $z<1.5$ than at earlier epochs.
However, the lack of detection of evolution could be due to
the inhomogeneity of the sample and/or to the
presence of some selection bias. 
The QSO obscuration bias may be responsible for missing
systems of higher  $Z/Z_\odot$
and higher column density (Section 3.1), which are expected
to give an important contribution to the metallicity at low redshift. 
The sample is inhomogeneous for studying the metallicity
evolution in the sense that  $Z/Z_\odot$ depends on the SFR and 
on $z_f$, two parameters that vary in different types of galaxies;
in addition $Z/Z_\odot$ can vary with the galactocentric
distance $r$ in a given galaxy.

Even if evolution is not directly detected from the data,
model predictions of metallicity evolution
obtained for different types of galaxies 
yield results consistent with the observations  
(Lindner, Fritze \& Fricke 1998; Jimenez et al. 1999). 
The models by Jimenez et al. (1999) show explicitely
the dependence of the metallicity on $z_f$ and on $r$. 
According to these models,  
LSB galaxies formed at $z_f=4$ (but not later)
fit well the DLA metallicities; HSB disks can account
for the observed data only if they form
continuously in the interval   $ 1 \leq z_f \leq 4 $.

The [$\alpha$/Fe] ratio is expected to decrease 
in the course of chemical evolution (Section 2.2). 
Measurements of the [Si/Fe] ratio are available for
almost 30 DLA systems (Lu et al. 1996; Prochaska \& Wolfe 1999), 
and for part of them
it is  possible to perform the  correction   for dust depletion 
(Vladilo 1998). 
The mean corrected value is not
significantly lower at $z<2$, where $<$[Si/Fe]$>=+0.01\pm 0.07 $ dex, 
than at $z>2$, where $<$[Si/Fe]$>=+0.10\pm 0.09$ dex (Fig. 1).   
The lack of evolution
could be due to the inhomogeneity of the
sample or to the presence of some selection bias, 
as in the case of the metallicity. 
In fact, all the selection effects 
that alter the study of the metallicities will also affect the
[$\alpha$/Fe] ratio, which evolves with metallicity. 
However, the [$\alpha$/Fe] ratios do not show evidence for evolution
even when plotted versus $Z/Z_\odot$ (Vladilo 1998).

\subsection{ Dust-to-gas ratios }

The dust-to-gas ratios $k$ estimated from the
iron-peak abundances do not show any  
trend with redshift. 
This is consistent with the fact that $k$
is very well correlated with $Z/Z_\odot$ (Fig. 2 in Vladilo 1998)
and the metallicity    does not   evolve with redshift.  

Since metallicity is an indicator of chemical evolution, 
the good correlation between $k$ and  $Z/Z_\odot$
can be considered as
evidence for evolution in DLA systems.
The regular increase of the dust content with metallicity, however, 
constrasts with the lack of any correlation with redshift. 
We deduce that DLA systems do evolve in a regular
fashion, but they attain a given level
of metallicity (or dust-to-gas ratio) at different redshifts.
This conclusion confirms that the sample of DLA systems  must
include galaxies with different
formation redshifts $z_f$ and/or different  SFRs.

\subsection{ Kinematics }
 
From the analysis of a sample of 16 absorbers,
Ledoux et al. (1998) find that the maximum $\Delta V$
at a given $z$ increases at lower redshifts. 
This result, if confirmed, would indicate that neutral 
regions exhibit increasingly faster motions with cosmic time.
However, analysis of the set of 28  measurements of $\Delta V$  
obtained by Prochaska \& Wolfe (1997, 1998)
does not confirm the existence of such a trend.  

Wolfe \& Prochaska (1998) find that
the maximum $\Delta V$ measured at a given
[Zn/H] increases with metallicity. 
According to these authors this effect can be 
explained by the passage of the lines of sight
through rotating   disks with radial
gradients in metallicity. 
An alternative explanation is that DLA galaxies 
with higher metallicities 
exhibit faster motions than DLA galaxies 
with lower metallicities. 
This interpretation would    fit nicely in a general
trend of increasing metallicity
with increasing mass (i.e., velocity dispersion).
In any case, the statistics are still insufficient to
confirm the existence of this trend.

\subsection{ Spin temperature }
 
As mentioned in Section 2.5,
the spin temperatures measured in high-$z$ DLA systems 
are higher than those measured in present-day spirals.
The difference
could be ascribed,  in principle, to an effect of evolution.  
However,  recent measurements  
in two DLA systems at $z_{\rm abs} =0.221$ and
$z_{\rm abs} =0.091$  yield
$<T_s> \, \approx 10^3 {\rm K}$, consistent with the values
found at high redshifts (Chengalur \& Kanekar 1999). 
This suggests that evolution may not be crucial and
confirms that DLA galaxies have properties
intrinsically different from those observed in nearby spirals.

\section{ Summary }
 
The nature of the galaxies hosting DLA clouds
is still a subject of debate. However, some important conclusions
can be inferred by comparing the results obtained
from different observations.   
At low redshifts,  candidate DLA galaxies in the fields
of the background QSOs show a variety of 
morphological types and different
levels of surface brightness. 
Spirals are not the dominant contributors, contrary to the
predictions based on the HI content of the nearby universe.
At high redshift, the hypothesis that DLA absorbers
originate in (proto)\-spirals is not supported
by spectroscopic studies of metallicities, abundance ratios,
dust-to-gas ratios and spin temperatures. 
In particular,  [$\alpha$/Fe] ratios and  nitrogen abundances   
hint at an origin in galaxies with properties similar to
those observed in  nearby, low-mass galaxies. 
Studies of kinematics are consistent
with an origin both in massive disks (proto-spirals)
and in low-mass galaxies.  

Evolution  effects are generally not detected in DLA systems.  
A possible exception is
the number density of $N_{\rm HI} > 10^{21}$ cm$^{-2}$ absorbers, 
which seems to decline at $z \geq$ 3.5.
The comoving mass density $\Omega_{\rm dla}(z)$ apparently
peaks at $z \approx 2.5$ and decreases at lower redshifts,
but this decrease 
is not corroborated by recent observations.
At the peak value $\Omega_{\rm dla}(z) < \Omega_{\rm star}(0)$,  
suggesting that we are missing part of the gas responsible
for the formation of present-day stars. 
Model predictions of LSB galaxies seem to better fit $\Omega_{\rm dla}(z)$
than models of HSB galaxies. 

Metallicities, abundances ratios, dust-to-gas ratios,
line-of-sight velocity dispersions, 
and spin temperatures do not show evidence of 
redshift evolution.  As a consequence,  
the differences between the properties of present-day spirals and
those of high-$z$ DLA systems cannot be ascribed to
evolutionary effects: DLA galaxies appear to be
intrinsically different from nearby spirals.

Dust production follows  metal production in a very regular
fashion in DLA systems. 
While this regular behaviour is  proof of evolution,
the lack of any correlation between metallicity
and redshift suggests that DLA galaxies attain a given 
level of metallicity at different cosmic epochs, i.e.
DLA  galaxies must have different
formation redshifts $z_f$ and/or different  SFRs.

Several selection effects conspire to bias the observed
population of DLA absorbers.
In particular, high column density
clouds located in environments with relatively high
metallicity can be missed  
owing to the QSO obscuration effect. This bias tends to
decrease the fraction
of spirals detected in the surveys. 
In general, the contribution
of the most chemically evolved galactic regions tends to be
underestimated.  On the other hand,
selection effects
tend to favour the detection of LSB galaxies  
and
the fraction of low-mass galaxies does not seem to be
underestimated.   
In spite of their small sizes,  low-mass galaxies  
can be detected if the
faint end of the luminosity function is sufficiently steep
(Tyson 1988),  a condition 
supported by results obtained at low $z$ 
(Zucca et al. 1997).

In conclusion, DLA absorbers appear to be associated with
a composite population of galaxies 
without strong effects of evolution 
and 
with a prominent representation of low-mass and/or LSB galaxies. 
A significant number of
massive galaxies can be detected in absorption
by increasing the statistics of   DLA surveys 
and  
by pushing the observational limits down to fainter magnitudes
in order to contrast selection effects.



\begin{references}

\reference
Boiss\'e, P., Le Brun, V., Bergeron, J., \& Deharveng, J.M.
1998, \aap, 333, 841

\reference
Braun, R., 1997, \apj, 484, 637  
 
\reference
Carilli, C.L., Lane, W., de Bruyn, A.G., Braun, R.,
Miley, G.K. 1996, \aj, 111, 1830

\reference 
Centuri\'on, M., Bonifacio, P., Molaro, P., \&
Vladilo, G. 1998, \apj, 509, 620

\reference
Chengalur, J.N., \& Kanekar, N. 1999, \mnras, 302, L29

\reference
de Bruyn, A.G., O'Dea, C.P., Baum, S.A. 1996, \aap, 305, 450

\reference
Dickey, J.M. 1995, in The Physics of the Interstellar and Intergalactic
Medium, eds. A. Ferrara et al., A.S.P. Con. Ser. Vol. 80, 357

\reference
Haehnelt, M.G., Steinmetz, M., \& Rauch, M. 1998, \apj, 495, 647

\reference
Jannuzi, B.T., Bahcall, J.N., Bergeron, J., Boksenberg, A., Hartig, G.F.,
Kirhakos, S., Sargent, W.L.W., Savage, B.D., Schneider, D.P.,
Turnshek, D.A., Weymann, R.J., \& Wolfe, A.M. 
1998, \apjs, 118, 1

\reference
Jimenez, R., Bowen, D.V., \& Matteucci, F. 1999,
\apjl, 514, L83

\reference
Kulkarni, V.P., S.M., Fall, \& J.W. Truran, 1997, \apjl, 484, L7

\reference
Lanzetta, K.M., Wolfe, A.M., Altan, H., Barcons, X., Chen, H.-W., 
Fern\'andez-Soto, A., Meyer, D.M., Ortiz-Gil, A., Savaglio, S., 
Webb, J.K., \& Yahata, N. 1997, \aj, 114, 1337

\reference
Lanzetta, K.M., Wolfe, A.M., \&
Turnshek, D.A. 1995, \apj, 440, 435


\reference
Lauroesch, J.T., Truran, J.W., Welty, D.E., \&
York, D.G. 1996, \pasp, 108, 641


\reference
Le Brun, V., Bergeron, J., Boiss\'e, P., \&
Deharveng, J.M. 1997, \aap, 321, 733

\reference
Ledoux, C., Petitjean, P., Bergeron, J., Wampler, E.J., \&
Srianand, R. 1998, \aap, 337, 51

\reference
Lindner, U., Fritze-Von Alvensleben, U., \& Fricke, K.J. 1998,
\aap, 341, 709
 

\reference
Lu L., Sargent W.L.W., Barlow T.A. 1998, \apj, 115, 55
 
\reference
Lu, L., Sargent, W.L.W., Barlow, T.A., Churchill, C.W., \&
Vogt, S. 1996, \apjs,  107, 475 



\reference
Matteucci, F. 1991,  in SN1987A and other Supernovae, ed.
I.J. Danziger \& K.Kj\"ar, ESO Proc. No. 37, 703

\reference
Matteucci, F., Molaro, P., \& Vladilo, G. 1997, \aap, 321, 45

\reference
Miller, E.D., Knezek, P.M., \&
Bregman, J.N. 1999, \apjl,  510, L95

\reference
Mo, H.J., Mao, S., \& White, S.D.M. 1999, \mnras, 304, 175

\reference
Molaro, P., Centuri\'on, M., \& Vladilo, G. 1998, \mnras, 293, L37

\reference
Molaro, P., D'Odorico, S., Fontana, A., Savaglio, S., \& Vladilo, G.
1996, \aap, 308, 1
 
\reference
Pei, Y.C., \& Fall, S.M. 1995, \apj, 454, 69

\reference
Pei, Y.C., Fall, S.M., \& Bechtold, J. 1991, \apj, 378,  6 
 
\reference
Pettini, M., Ellison, S.L., Steidel, C.C., \& Bowen, D.V. 1999,
\apj, 510, 576

\reference
Pettini, M., King, D.L., Smith, L.J., \&
Hunstead, R.W.  1997a, \apj, 478, 536

\reference
Pettini, M.,  Smith, L.J., King, D.L., \&
Hunstead, R.W.  1997b, \apj, 486, 665


\reference
Prochaska, J.X., \& Wolfe, A.M. 1997, \apj, 474, 140

\reference
Prochaska, J.X., \&  Wolfe A.M. 1998, \apj, 507, 113

\reference
Prochaska, J.X., \&  Wolfe A.M. 1999, \apj, in press (astro-ph/9810381)

\reference
Rao, S.M., \& Briggs, F. 1993, \apj, 419, 515

\reference
Rao, S.M., \& Turnshek, D.A. 1998, \apjl, 500, L115

\reference
Ryan, S.G., Norris, J.E., \& Beers, T.C. 1996, \apj, 471, 254


\reference
Savage B.D., \& Sembach K.R. 1996, Ann. Rev. Astron. Astrophys., 34, 279

\reference
Smette, A., Claeskens, J.-F., \& Surdej, J., 1997,
New Astronomy 2, 53
 
\reference
Steidel, C.C., Dickinson, M., Meyer, D.M., Adelberger, K.L., \&
Sembach, K.R. 1997, \apj, 480, 568


\reference
Storrie-Lombardi, L.J., Irwin, M.J., \& McMahon, R.G. 1996b, \mnras,
282, 1330 

\reference
Storrie-Lombardi, L.J., McMahon, R.G., \& Irwin, M.J.  1996c, \mnras,
283, L79  

\reference
Storrie-Lombardi, L.J., McMahon, R.G., Irwin, M.J., \&
Hazard, C.  1996a, \apj, 468, 121

\reference
Thuan, T.X., Izotov, Y.I., \& Lipovetsky, V.A. 1995, \apj, 445, 108
 
\reference
Turnshek, D.A. 1998, in Structure and Evolution of the Intergalactic Medium
from QSO absorption Line Systems, ed. P. Petitjean, \& S. Charlot
(Paris:Editions Frontieres), 263

\reference
Tyson, N.D. 1988, \apjl, 329, L57

\reference
van Zee, L., Haynes, P.M., Salzer, J.J., \& Broeils, A. 1996,
\aj, 112, 129

\reference
Vladilo, G. 1998, \apj, 493, 583

\reference
Vladilo, G., Centuri\'on, M., Falomo, R., \&
Molaro, P. 1997, \aap, 327, 47

\reference
Vladilo, G., Molaro, P., \& Centuri\'on, M. 1999,
Proc. of the meeting The Birth of Galaxies,  Chateau de Blois, France, 
June  28th - July  4th (1998), 
in press.  

\reference
Wheeler, J.C., Sneden, C., \& Truran, J.W.Jr. 1989,
Ann. Rev. Astron. Astrophys., 27, 279

\reference 
Wolfe, A.M., Prochaska, J.X. 1998, \apjl, 494, L15


\reference
Wolfe, A.M., Lanzetta, K.M., Foltz, C.B., \&
Chaffee, F.H. 1995, \apj, 454, 698

\reference
Wolfe, A.M., Turnshek, D.A., Smith, H.E., \&
Cohen, R.D. 1986, \apjs, 61, 249


\reference 
Zucca, E., Zamorani, G.,  Vettolani, G., Cappi, A., 
Merighi, R., Mignoli, M., Stirpe, G. M., 
       MacGillivray, H., Collins, C., Balkowski, C., Cayatte, V., 
       Maurogordato, S., Proust, D., Chincarini, G., 
       Guzzo, L., Maccagni, D., Scaramella, R.,  Blanchard, A., \&
       Ramella, M. 1997, \aap, 326, 477
       
\reference 
Zuo, L., Beaver, E.A., Burbidge, E.M., Cohen, R.S., Junkkarinen, V.T., \&
Lyons, R.W., 1997, \apj, 477, 568


\end{references}
\end{document}